%% file: CSCW_Digital_Traces_paper.tex
\documentclass[authorversion]{acmart}

\usepackage{multirow}
\newcommand{\anon}[1]{\textcolor{black}{~(anon.)}}

\AtBeginDocument{%
  \providecommand\BibTeX{{%
    \normalfont B\kern-0.5em{\scshape i\kern-0.25em b}\kern-0.8em\TeX}}}

\setcopyright{acmcopyright}
\copyrightyear{2022}
\acmYear{2022}
\acmDOI{}

\acmConference[CSCW '22]{CSCW '22:THE 25TH ACM CONFERENCE ON COMPUTER-SUPPORTED COOPERATIVE WORK AND SOCIAL COMPUTING}{Nov 12--16, 2022}{Taipei}

\begin{document}

\title{Revealing Cumulative Risks in Online Personal Information: A Data Narrative Study}


  
\author{Emma Nicol}
\affiliation{
\institution{University of Strathclyde}
\country{UK}}
\author{Jo Briggs}
\affiliation{
 \institution{Northumbria University}
  \country{UK}}
\author{Wendy Moncur}
\affiliation{
 \institution{University of Strathclyde}
\country{UK}}
\author{Amal Htait}
\affiliation{
\institution{Aston University}
\country{UK}}
\author{Daniel Carey}
\affiliation{
\institution{Independent Researcher}
\country{UK}}
\author{Leif Azzopardi}
\affiliation{
\institution{University of Strathclyde}
\country{UK}}
\author{Burkhard Schafer}
\affiliation{
\institution{University of Edinburgh}
\country{UK}}

\begin{abstract}
  
   When pieces from an individual’s personal information available online are connected over time and across multiple platforms, this more complete digital trace can give unintended insights into their life and opinions. In a data narrative interview study with 26 currently employed participants, we examined risks and harms to individuals and employers when others joined the dots between their online information. We discuss the themes of \textit{visibility and self-disclosure}, \textit{unintentional information leakage} and \textit{digital privacy literacies} constructed from our analysis. We contribute insights not only into people's difficulties in recalling and conceptualising their digital traces but of subsequently envisioning how their online information may be combined, or (re)identified across their traces and address a current gap in research by showing that awareness is lacking around the potential for personal information to be correlated by and made coherent to/by others, posing risks to individuals, employers, and even the state. We touch on inequalities of privacy, freedom and legitimacy that exist for different groups with regard to what they make (or feel compelled to make) available online and we contribute to current methodological work on the use of sketching to support visual sense making in data narrative interviews. We conclude by discussing the need for interventions that support personal reflection on the potential visibility of combined digital traces to spotlight hidden vulnerabilities, and promote more proactive action about what is shared and not shared online. 
\end{abstract}

\maketitle

\input{introduction}

\input{related_work}
\input{methodology}

\input{results}

\input{conclusions}

\begin{acks}
We acknowledge the contributions of our project partners and time and effort of our participants. This work was sponsored by EPSRC grant EP/R033889/1. The study was approved by the Research Ethics Committee of DJCAD, University of Dundee. 
\end{acks}
\bibliographystyle{ACM-Reference-Format}
\bibliography{references}

\clearpage\section{Appendices}
\appendix

\section{Interview Questions}
 
\begin{table}[H]
  \caption{Interview Questions with Changes in Response to Lockdown/ Working From Home}
  \label{tab:freq}
  \begin{small}
  \begin{tabular}{lcl}
    \toprule
    Question & Original(Y/N) & Amended (Am)?\\
    \midrule
    What communication channels do you use?&Y&Am\\ 
    What Apps and Sites (e.g Twitter, match.com, Strava) do you use? & Y & Am\\
    What data management services do you use (e.g. iCloud, Box)? & Y & Am\\
    Devices used (e.g. mobile, Fitbit, NEST, ioT) & Y & Am - sharing?\\
    Do you share devices with anyone? & Y & Am - emphasis\\
    Behavioural patterns/practices (e.g. search history) & Y & Am\\
    What do you post? Who it is visible to?& Y & Am\\
    Types of information posted, Why/how do you hide data & Y & Am\\
    Instances where information posted revealed more than intended & Y & Am\\
    Concerns about types of revelation & Y & Am\\
    What are your digital hygiene practices & & \\
    - to ensure privacy/reputation mgmt? & Y & Am\\
    Account sharing and associated posting responsibilities & Y & Am\\
    Given an overview of your data, what could be found out? & Y & Unchanged\\
    Do you work in public/private sector? What level? How big is org?& N & Added\\
    Are you working from home during Lockdown? Is this new? & N & Added\\
    Did your employer advise on data management during Lockdown?    & N & Added\\
    Who lives at home with you? Any bandwidth issues? & N & Added\\
    What advice would you give to someone with & & \\ 
    the same eco-system as you, to be secure? & N & Added\\
 \bottomrule
\end{tabular}
\end{small}
\end{table}

\clearpage\section{Survey Outcomes Overview}

\begin{figure}[H]
    \includegraphics[width=\linewidth]{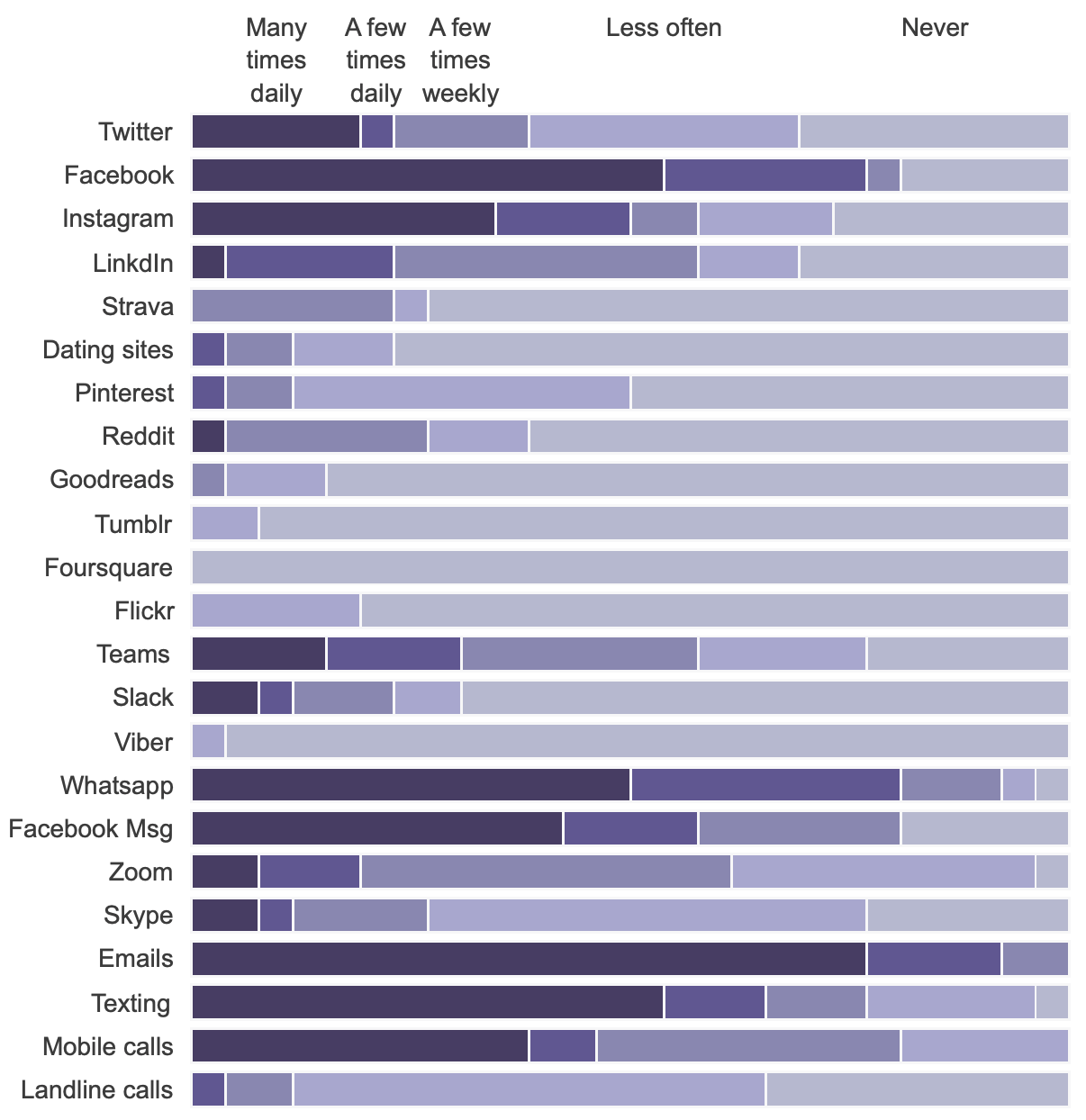}
    \caption{Participant Responses to Technology Questionnaire. 
    }
    \label{fig:participant_responses}
\end{figure}

\clearpage\section{Participant Sketches}

\begin{figure}[H]
    \centering
    \includegraphics[width=0.485\columnwidth]{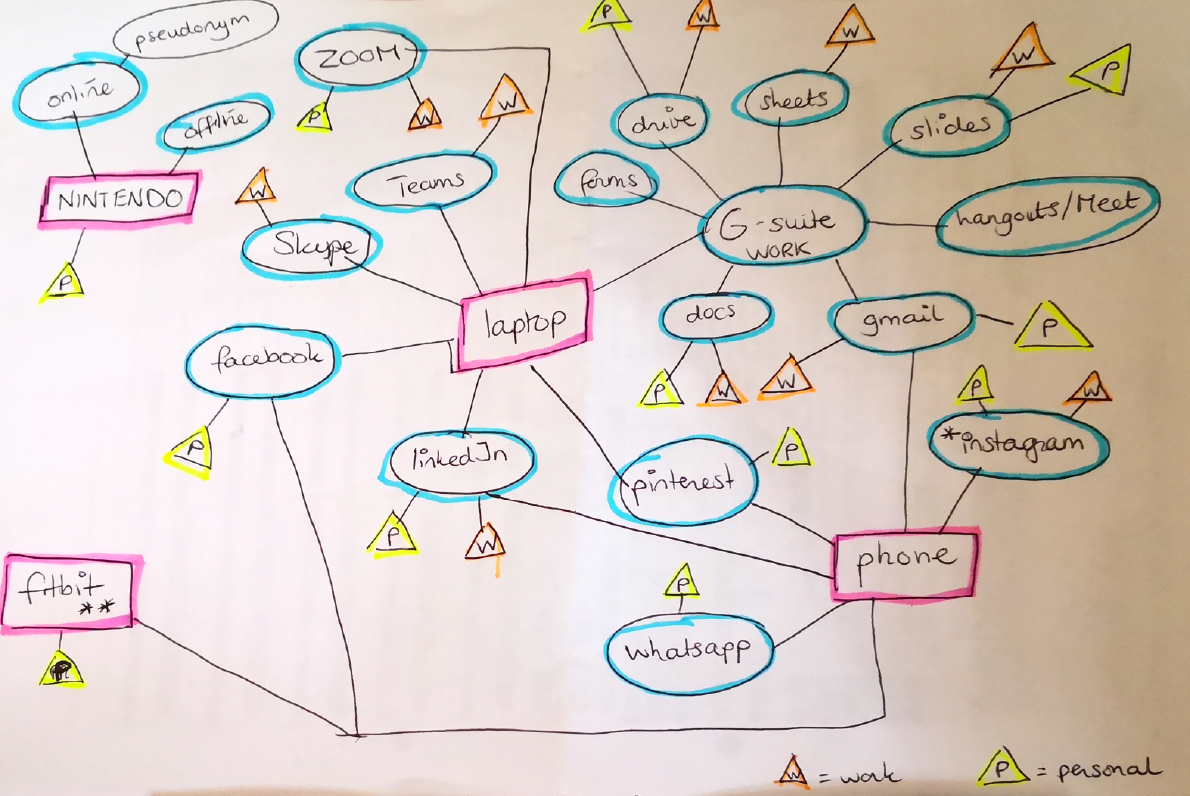}
    \includegraphics[width=0.485\columnwidth]{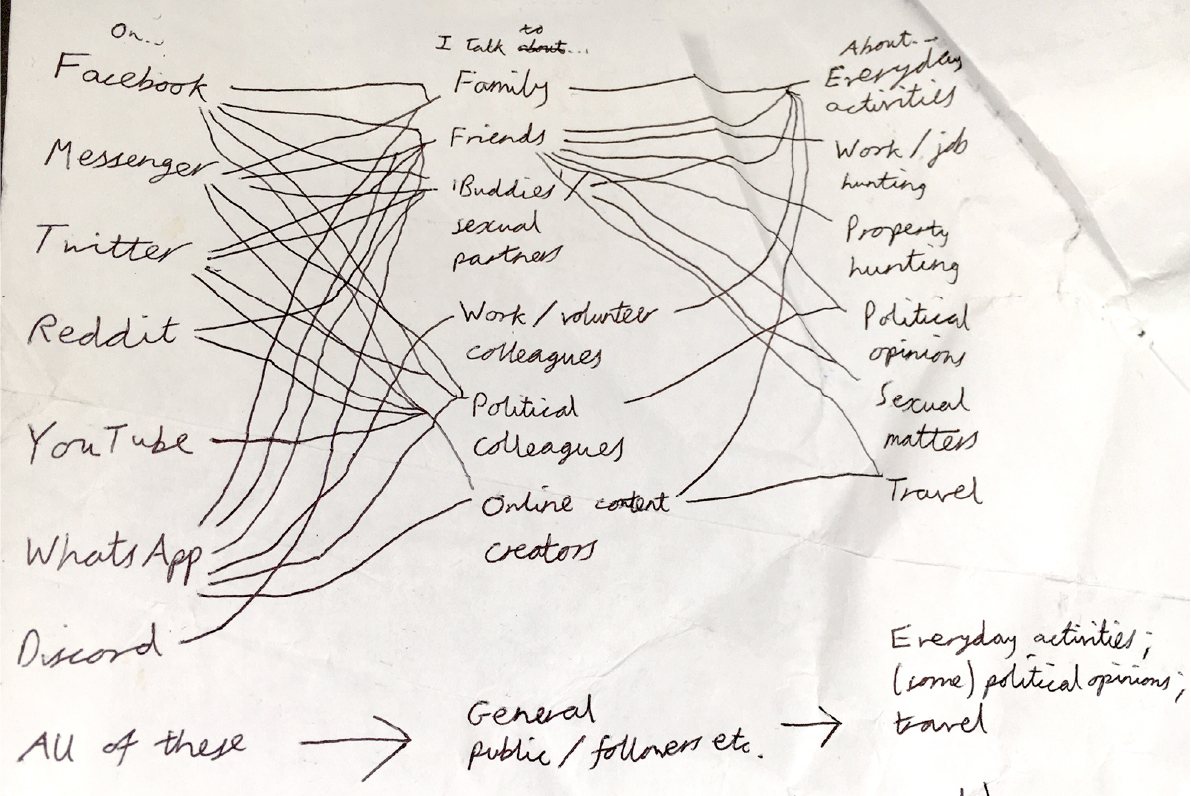}
    \caption{Two drawings created by participants, Vinny, 24 (right) and Denise, 32 (left) during interview to represent the online platforms they used and the types of information shared. Both participants indicated interconnections, drawing lines between platforms to convey information flows. We received sketches from 21 participants documenting their devices and apps, often along with the various relationships these enabled, in a mapping or table arrangement. The sketches supported and added depth to interview conversations and the drawing of the sketches often triggered participants to remember additional details }
 \label{fig:participant_drawings}
\end{figure}

\end{document}

%% file: introduction.tex
\section{Introduction}\label{sec_introduction}

Our connections to diverse online services through multiple personal devices make it increasingly difficult to keep track of what we are disclosing about ourselves online via our digital traces. These traces include those that arise from people sharing their own personal information online, others sharing information about them, and via e.g., automated functions that make additional metadata public, such as disclosing one's location when posting on Instagram. Sharing may be intentional, unintentional or inadvertent. People are connecting digitally with others for professional and social reasons, via an increasing diversity of channels. Such diversity means that networks supporting interaction and collaboration with others are evolving in ways that are complex and difficult to mentally model. The challenge of understanding how one's data is being shared in and across these networks grows with this complexity. 

A substantial body of research has examined how an individual’s digital traces may be used to discover or infer information about them --- their interests, livelihood, place of work, whether they are depressed or likely to self-harm, relationships, sexual orientation, political opinions, religion and other preferences --- even when not explicitly disclosed ~\cite{Krasnova2010,Xia2013,Haimson2016}. Smartphone use alone can reveal much, based on information including accelerometer and GPS data, app usage patterns, call logs and Bluetooth proximity \cite{Ellis2016TheThreats}. Revelations include the phone user’s identity, mood, stress levels, personality, whether they are a parent, likely destination when travelling, whether they are sitting/walking or running, and the quality of their sleep \cite{Ellis2016TheThreats}. 
When combined over time and across multiple digital channels --- e.g., Facebook, LinkedIn and Tinder --- this array of digital traces can afford unintended insights into people’s lives as private individuals, employees, and citizens. Combined digital traces may also reveal unintended insights about employers, and even about national security. 

These insights are significant --- from them, we can learn just how much we are revealing about ourselves online, not just about our past and present, but also about our future: {“...fragments of past (online) interactions or activities …, when correlated together, allow a preemption and prediction of future behaviors”} \cite[p250]{Reigeluth2014WhySelf-control}. Insights are also relevant to hostile actors --- e.g., fraudsters --- who can make use of coherent, combined digital traces to gain advantage over their victims. 

 In this research we investigate:
 
 \begin{itemize}
  \item The everyday online information sharing practices of employed people, and their associated awareness of how pieces of personal information --- digital traces --- can be connected over different online channels over time;
  \item To what extent people recognise how their connected traces are available to others, potentially to be explored as a more coherent whole;
  \item What this more coherent whole could convey about an individual, including insights into their apparently private self (e.g., behavior, values, habits etc.);
  \item  How aware people are of the potential harms and hazards of how such insights could be used against them, and by whom.
\end{itemize}

 We took a holistic approach, acknowledging the interconnectedness of online practices ~\cite{Madianou2012Polymedia:Communication:,vertesi_data_2016-1}; the networked nature of online identity~\cite{Marwick2014NetworkedMedia}; and use of multiple digital channels (e.g., social networking sites, IoT, apps), devices (e.g., personal smartphones, Wi-Fi-enabled devices at home, work computers), and behavioral patterns that are determined by the affordances of particular digital technologies (e.g., GPS data of fitness apps) \cite{Ellis2019DoBehavior}. We conducted 26 semi-structured interviews to (i) solicit insights into interviewees' digital ecosystems across multiple communication channels, sharing networks and devices plus associated behavioral patterns and practices; (ii) explore co-constructed aspects of participants' online identities across apparently discrete channels of information and; (iii) identify experiences and consequences where combined digital traces revealed more than intended. 

All participants in this study were residents of the United Kingdom (UK), and subject to the protection of the EU General Data Protection Regulation (GDPR) in tandem with the UK Data Protection Act 2018 (DPA) \cite{Lynskey2015TheLaw}. However, following the 2016 referendum at which the UK decided to leave the European Union, the GDPR stopped being directly applicable. We were mindful of this uncertain context while conducting the study and we therefore acknowledge the implications for our participants and findings in our discussion.

We first situate our study in the context of prior work on digital traces, cybersecurity, and the workplace, before going on to outline our data narrative interview method, accounting for the necessary changes to its design during Covid-19 related Lockdown conditions. We then present the results of our thematic analysis across three themes: \textit{visibility and self-disclosure}, \textit{unintentional information leakage} and \textit{digital privacy literacies}. We discuss issues around soliciting people's recollections and understandings of their digital traces across networked space and time. Our contribution comprises insights not only into people's difficulties in recalling and conceptualising their digital traces but of subsequently envisioning how their information may be combined, or (re)identified across their traces. We also contribute insights on the inequalities of privacy, freedom, and legitimacy that exist for different groups with regard to what they make or feel compelled to make public online and the privileges that are enjoyed by some but not by others. Finally, we make a methodological contribution on the use of sketching to support visual sense making in the interviews, inviting new perspectives on researching personal online information interactions, and building on previous studies in CSCW that used this method in other contexts. This set of interviews represents a step towards our overarching research goal to identify the need for tools or other designed interventions that support not only personal reflection on the potential visibility of combined digital traces, but that additionally support the curation of one's existing traces, retrospectively.

%% file: related_work.tex
\section{Related Work}\label{sec_related_work}

Our study takes a socio-technical approach to the cybersecurity risks emanating from people’s digital traces. We interpret cybersecurity from a post-digital perspective, where ``the protection of technology and information has become so intermingled with the protection of people and society that distinguishing between the two is impossible. …in a post-digital society, technological security rests upon the protection of people, and vice versa'' \cite[p10]{Coles-Kemp2020TooSociety}. Prior work by Dunphy et al. \cite{Dunphy2014UnderstandingTechnologies} at the intersection of design and cybersecurity has focused on people and their experiences, while government agencies such as the UK’s National Cyber Security Centre (NCSC) have now introduced cybersecurity guidance ``for anyone looking to develop security which works for organisations and for people'' incorporating a design orientation \cite{ncsc_cyber_2018-1}. Our specific focus is on combined, intersecting digital traces, and we frame our work against this backdrop, first defining digital traces for the purposes of this paper and explaining how they can be (mis)used, and the digital literacies that affect people’s understanding of how they `look' online.

\subsection{Defining Digital Traces} 
Our digital traces are multi-dimensional. We leave traces of personal information across multiple digital platforms and across time. Such traces are generated ~\cite{Reigeluth2014WhySelf-control} before we are born, across the lifespan~\cite{Orzech2018OpportunitiesUK}, and even post-mortem \cite{Moncur2016LivingDigitally}. Even in childhood, there is a multiplicity of apparently innocuous channels via which personal information is often shared --- for example through connected toys; children and parents posting on social media; biometrics used for schools' fingerprint charge accounts ~\cite{UKChildrensCommissioner2018WhoMe}. Brandtzaeg and L\"{u}ders~\cite [p2]{Brandtzaeg2018TimeCollapse} highlight that it is ``increasingly important to understand how \textit{time} is perceived in the context of a non-anonymous social media environment'' (authors' emphasis), as digital traces over time can reveal much, not only about our current selves, but also about our past opinions, actions and feelings.

Digital traces emanate from the central actor, using their real name and pseudonyms. They also emanate from a range of other actors --- e.g., health providers; employers (including through productivity tracking \cite{Bannerman2019RelationalSelf}); government agencies (including public registers of companies) --- serving to produce a co-constructed digital identity for individuals \cite{Tufekci2008GROOMINGMYSPACE}. In addition, people's personal information can surface through other channels, posted by friends and acquaintances, and also by government sources (e.g., voter registers) and other organisations (e.g., that collate and create dossiers on individuals). 

While these traces are spread across multiple locations, with subsets of information shared with specific audiences, boundaries between the platforms and audiences are known to be porous. Context collapse, ``in which individuals must meet the expectations of multiple and diverse audiences simultaneously'' \cite [p2]{Brandtzaeg2018TimeCollapse} is a recognised phenomenon. Embarrassing and harmful situations can arise through context collapse, when users try to navigate multiple, diverse audiences on the same platform: they may accidentally blur borders between the public and the private, the professional and the personal --- leading to information leakage  \cite{Davis2014ContextCollisions}. However, Costa \cite [p3652]{Costa2018Affordances-in-practice:Collapse.} found that context collapse was not a given, and was ``a result of situated practices of social media usage within Western Anglophone contexts”. Her Turkish participants ably navigated complex security settings to ensure that boundaries between online groups were maintained. 
\subsection{Use of Combined Digital Traces}
There are tools available that enable individuals to make sense of their digital traces in certain contexts. For example, quantified self, personal informatics, and life logging tools can give people a better understanding of their own behaviors, through recording, measurement, visualization and publication of their own data \cite{Elsden2016AInformatics, Sellen2010BeyondLifelogging,  Li2010ASystems,Reigeluth2014WhySelf-control}. Coherently combined traces --- e.g. geolocation, step count, heart rate --- can become material for conversation and expression of personal identity, and/or to improve behavior or performance in a particular area, and form ``highly personal accounts of (users) pasts'' ~\cite[p518]{Elsden2016AInformatics}. The utility of combined digital traces extends beyond the individual: exploitation by others can afford unintended insights and privacy violations, potentially adversely affecting individuals, employers and organisational security. 

Approaches to exploiting digital traces may be manual or involve the use of specially-developed tools. 'Lurkers' --- especially adolescents ---  may trawl the social media feeds of friends and followers for updates and juicy titbits, joining the dots between posts to work out more than was intended to be revealed \footnote{https://psycnet.apa.org/fulltext/2017-07146-004.pdf}. More seriously, perpetrators of intimate partner violence may go to great lengths online to track down their victims (survivors) through their digital traces across multiple platforms, in order to continue their abusive behavior \cite{Grimani2020AnViolence}. Examples of tools developed to harvest digital traces include a Blockchain-based application that enabled people to establish others' trustworthiness \cite{Yang2020TAPESTRY:Online}, and a tool that combined online dating site posts with fitness tracker information to reveal where people lived, whether they lived alone, and when they were at home or out exercising --- information that was subsequently exploited by stalkers ~\cite{Chen2016}. Using Facebook profiles, \citet{Bachrach2012FacebookPersonality_michal_29_04_12.pdf} were able to infer people's Big Five Personality traits. Other initiatives have sought to predict from Twitter and Reddit posts whether people are depressed, suffering from anorexia, or likely to self-harm ~\cite{Losada2019}. On a larger scale, and beyond our current focus on an individual's interpretation of combined digital traces, automated Big Data approaches can use these same traces to assemble insights into people’s behavior and to predict e.g., the likelihood of someone repaying a loan or developing diabetes ~\cite{Pentland2012ReinventingData}, someone's political leanings ~\cite{OpenRightsGroupWhoGroup}, their propensity for criminal behavior \cite{Meijer2019PredictiveDrawbacks}, and their mental health  \cite{Losada2019}. 

Efforts to exploit digital traces can be facilitated via lax security and privacy settings and behaviors. A large scale example of this lies in the lax security applied to social networking sites in Runet (the Russian Segment of  the Internet), leading to 30 million profiles becoming publicly available to download ~\cite[p50]{Kisilevich2010AnalysisRunet}. The profiles included users' personal and intimate details such as ``sexual orientation, sex frequency and preferences in sex'', along with ``personal information like weight, height, smoking habits, alcohol, drugs, body characteristics... dwelling type, marital status, and religion''. While this information may have been disclosed to circulate within the particular context of e.g., a dating site, it became more widely available, exposing people's intimate details to a wider audience than was ever intended ~\cite{Kisilevich2010AnalysisRunet}. On a smaller scale, when parents share posts and pictures of their children (``sharenting'') without consent, they add a dimension to their children's online identities that may be at odds with how their children wish to appear online, skewing their digital traces ~\cite{Ouvrein2019Sharenting:Management}. Meanwhile sharing posts and photos about friends may unwittingly reveal private information about them (e.g., details of companions or associates, locations, activities etc). 
\newline
Even when people actively separate their digital identities across different online channels to organize their social groups or to obfuscate aspects of their identity, their efforts can be undermined by re-identification, involving the linking together of profiles and other information  ~\cite{Gross2005InformationNetworks}. For example, the Personal Genome Project ~\footnote{\url{https://www.personalgenomes.org/}} linked profiles to online voter lists via e.g., birthdate, zip code --- destroying the assumed anonymity of the profiles ~\cite{Sweeney2013IdentifyingExperiment}; Facebook profile images tagged with real names were used to re-identify people on other sites (e.g., Friendster, Match.com) that host otherwise anonymous profiles~\cite{Gross2005InformationNetworks}.


\subsection{Employees, Employers and Organisations}
Social media content is used by up to 80 percent of employers and recruitment agencies as part of their assessment of candidate suitability for a post~\cite{Sprague2011InvasionRelationship}. ~\citet{Young2009} found that applicants’ chances of acceptance for advertised job positions could be adversely impacted by their openness online about e.g., health conditions or pregnancy. Employees can face dismissal for their conduct on social networking sites, even when posting outside of working hours \cite{Thornthwaite2018SocialPrivacy}. \citet[p119]{Thornthwaite2018SocialPrivacy} observes that social media is {``blurring the legal distinction between employees’ public and private lives, increasing employer control over personal lives in ways reminiscent of traditional master–servant relationships''}, the effects of which are then tested in industrial tribunals, which increasingly challenge employers' intrusive stance. 

Employees' social media activities also have the potential to negatively impact on their employers by unintentionally leaking sensitive information online --- such as trade secrets, intellectual property and personal details of other employees. This can represent a significant security risk to organisations {``result(ing) in a loss of competitive advantage, loss of reputation, and erosion of client trust''}~\cite[p351]{AbdulMolok2018ANetworking}. Irresponsible posting can result in damage to employers that goes way beyond their reputation. There have been instances of military personnel and their families discussing operations and deployment details on social media, even posting pictures of ongoing operations ~\cite{Dressler2015}. In a recent case, staff living on a UK nuclear submarine base exposed compromising information via their use of the Only Fans pornography-sharing website ~\cite{Metro2021NavyNews}. Adversaries motivated to exploit such information can seek out service personnel or their family members to blackmail them, or use the intelligence gathered to attack or infiltrate deployment locations. 

\subsection{Digital Self and Contextual Privacy}

The work of boyd e.g., \cite{Boyd2008TakenPublics}\cite{Boyd2009WhyLife}     \cite{Boyd2010PublicScholar} and boyd and Ellison \cite{Boyd2007SocialScholarship} drew from Goffman's notions of impression management \cite{Goffman1959TheLife} to articulate the necessary maintenance and updating of ``front stage'' digital selves and online identities, and discrepancies between what one ``gives'' --- in explicit displays of friends, interests and online representations --- and ``gives off'' as interpreted by others, and as  amplified in networked publics \cite{Boyd2008TakenPublics}\cite{Boyd2009WhyLife}. boyd's early work with young people also found that they cared deeply about online privacy, and recognised the value of both privacy and publicity, understanding them as contextual and adjustable, the latter especially with regard to being socially present and acquiring social status. Indeed, self-disclosure is a key factor in developing relationships and building trust in online environments, much as it is in face-to-face contexts. This  self-disclosure includes actions such as deliberately sharing selected personal photographs \cite{Dindia2000SexReviewed}, and represents a negotiation of privacy, with information often shared with selected (groups of) online network members, rather than with all network members \cite{Marwick2014NetworkedMedia}.


This crafted, contextually-situated privacy is subject to the pressures of normativity. When flows of information adhere to entrenched norms, there are few concerns, whereas when violations of norms occur, protest and complaint often result \cite{Nissenbaum2011AOnline}. For example, a patient may be comfortable with healthcare providers sharing medical information with specialists, but very uncomfortable when the same data is shared with marketing companies (see also \citet{Bussone2020TrustHIV}). When considering contextual privacy in workplace or organisational settings, we can look to the work of e.g., \citet{Ashenden2018InSecurity}. This work found that employees who believed their employer was driven by the need to protect information thought risks to be overstated and colleagues overly cautious, whereas those who believed the organisation was driven by a need to optimise information use, thought security risks justified and colleagues’ behavior risky.
 \citet{McDonald2020TheVulnerabilities} have drawn attention to the limits of concepts such as contextual integrity and boundary regulation when thinking about privacy in Human Computer Interaction (HCI). They revealed conceptual gaps in current frameworks and have argued for considering vulnerability i.e. of particular groups of people as a core concept when thinking about privacy. Researchers in Computer Supported Cooperative Work (CSCW) have highlighted serious issues with digitally-mediated identity management, with the effects often being pronounced for those considered vulnerable e.g., Simpson and Semaan’s work on algorithmic identity investigated and confirmed concerns that the short video sharing application TikTok was suppressing the individual identities of LGBTQ+ users via algorithmic and human moderation \cite{Simpson2021ForForYou}.
Seberger et al. \cite{Seberger2021EmpoweringThat} showed how users navigate trade-offs involved in app use. Despite technical and regulatory mechanisms aimed at empowering users to manage their privacy, people have a sense of resignation around privacy due to the convenience offered by apps: there is a fine line between feeling empowered by technology and the discomfort of invasive app behavior. Users are often resigned to disclosing data even as they accept personal responsibility for their own privacy.

\subsection{Data Leakage}
People can often be surprised when they discover the personal data collection and distribution activities of apps that they use. Shklovski et al. \cite{Shklovski2014LeakinessUse} showed that people felt personal space had been violated in ``creepy'' ways by apps with the creepiness lying in the realization that apps were conduits for personal information and space to ``leak'' to unknown entities who had not explicitly been invited in. These authors link creepiness to notions of personal space and territoriality \cite{Altman1975TheCrowding.}, and to contextual integrity \cite{Nissenbaum2009PrivacyLife}. People will formally agree to the information sharing undertaken by apps, and can rationalize their use of them when asked, putting any creepiness out of their minds. Nonetheless, the creepiness remains. \cite{Shklovski2014LeakinessUse} further suggest that there are harmful health consequences of this enduring creepiness while acknowledging that such creepy experiences may not always be negative and unwanted; and pose an interesting question as to whether such creepiness fades over time suggesting that as cultural norms change, so too will the conditions under which such creepy experiences are encountered, which has implications across the digital lifespan.

\subsection{Digital Privacy Literacies} 
A person's ability to generate digital traces online does not imply the accompanying presence of digital privacy literacy, that is, an understanding of how information travels when shared online, and the associated risks. Digital privacy literacy is an area of education and practice often used in social justice and/or public education programmes, including those run by public libraries to help develop competencies in groups of people at particular risk, including those who rely on public library computers for personal digital communications and information practices, particularly those who are subject to social inequalities (see e.g.,  ~\cite{2015DataLibrarians}). Digital privacy literacy can be considered distinct from the well-established, if rather broad area of \textit{digital literacy} (e.g., ~\cite{Kress2003LiteracyBooks}), which refers to an individual's multiple competencies, from having regular access to and basic functional operational skills (e.g., using a keyboard and mouse) along with being able to ``read'' and ``write'' clear information across a number of digital modes using these apparatus, including in textual, visual and wider forms of communication media \cite{Kress2003LiteracyBooks}. Digital privacy literacy can be regarded as a subset of \textit{data literacy}, and is used to refer to a range of competencies around understanding and communicating with informational data, which often relates to privacy and (personal) data, and goes towards enabling one's personal data self-care.

In the social sciences, the growing literature on critical data studies e.g., \cite{Kitchin2015SmallData}, includes \cite{Lupton2019} Lupton's notion of the ``lively'' aspects of personal data, as they are added to, and (re)configured by human interpretation and corporate segmentation/analyses. Lupton notes that personal data, as (partly) human, is typically represented visually and in language  as organic and material (e.g., flows, breadcrumbs) or ``humanised'' as e.g., footprints (p47). People’s encounters with the personal data that they generate via use of digital technologies presents them with challenges as to how to interpret, control and make sense of these data. Lupton elsewhere \cite{Lupton2017FeelingData:} has argued that such data and their circulations could be made more perceptible and interpretable using what she describes as three dimensional materialisations, recognising that people’s interactions with such re-presentations of personal data elicit visceral responses. There are also new research areas around human data interaction that include designing frictions into user experiences of technologies to promote critical reflection, e.g., prior to sharing information; and on interaction design's dark patterns, which comprise e.g., targeted manipulation and confusing terms of service \cite{Gray2019Whendark}, widely adopted by industry to nudge people into a particular course of action, including coercing people into disclosing personal data, often beyond that necessary for the task in hand. 

%% file: methodology.tex
\section{Methodology}\label{sec_methodology}

We conducted an interview study using a data narrative approach \cite{vertesi_data_2016-1} in order to understand the risks and consequences of the digital traces that people leave online. This approach served to capture participants’ descriptions of their data, device use, channels and networks of communication, and data and information practices. Using this approach also allowed us to capture the co-constructed aspects of a person's online identity, as well as enabling the investigation of direct and observed experiences of, in this case, the cumulative implications of digital traces. 

\subsection {Interview Study}
We conducted the study in May-July 2020. Due to the physical distancing requirements of the Covid-19 Lockdown in place in the UK at the time of the study, we had to conduct interviews remotely via videoconferencing, which we took care to pilot before the interviews. We engaged with participants in advance by sending information sheets by email. Cognisant of the likely effects on data sharing that might arise due to the circumstances of the Lockdown, we added questions to the interview schedule regarding changes to data-sharing habits and experiences, with the intention of capturing the effects of self-isolation, homeworking and other Lockdown-related phenomena (Appendix A). 

\subsection{Participants} 
We recruited 26 adults (13 male, 12 female, one non-binary; age 20 -- 59 years, median 37 years) to take part in the study. All were based in the UK, active online and in full-time employment. We recruited participants by creating an advertisement that was circulated via emails to contacts with access to mailing lists to be shared more widely, and via social media, with the offer of a £20 shopping voucher for participation. We aimed to recruit participants from a variety of employment roles and sectors and made sure there was roughly equal representation from employees in the public (n=16) and private sectors (n=10) and that staff recruited represented all levels of seniority. Participants were employed in sectors including healthcare, education, engineering, management, IT and hospitality and were drawn from city, suburban, town and rural locations. 21 participants reported speaking English as their first language, four spoke it as a second language and one was bilingual in English and another language. 13 had postgraduate qualifications, 10 were qualified to undergraduate level, two had qualifications from further educational studies and one had high school qualifications. When we asked about their level of technology skill (the interviewer read out the full definition of each category to each participant) they responded as follows: four said {``Low/Low-Medium''} indicating basic use of software, hardware and social media; 15 said {``Medium''} indicating confidence with using and integrating a variety of standard software packages over a number of platforms; seven said {``Medium-High/High''} including the use of specialised software and an ability to program. At the time of interview, 18 participants were working at home full time, five split their time between working at home and at their workplace, while three reported no home working.

\subsection{Method}
The first author conducted all interviews, with each lasting 60-90 minutes. In a short briefing, we invited participants to ask questions about the study based on their reading of the information sheet. They then provided consent verbally. We then asked participants to complete a technology questionnaire via a SurveyMonkey link, delivered via the Chat function of the videoconferencing software, or sent by email. We designed the questionnaire to capture participants' self-reported use of technology including devices, communication channels, data storage and social media networks (Appendix B). In addition, the interviewer asked participants a short series of questions regarding their current employment, level of education, and their understanding of and confidence in using digital technology.

Interview questions were centred on the following areas and are detailed in Appendix A: (i) information about communications channels, apps, data storage/management systems, and devices used, including whether/how any of these were shared; (ii) everyday practices and behavior patterns around e.g., conducting searches, posting and other digital information sharing; (iii) participants’ awareness of the unanticipated potential for self-disclosure through digital traces and their associated level of concern; iv) information management, security setting behaviors and Lockdown-related changes --- especially regarding working from home; v) we asked participants to envision a scenario where someone else had to write a book about them based only on their digital traces, and to think about what the resulting book would comprise. We finished by asking them to summarise their advice to others on optimising their information security. We tailored the questions, where appropriate, through answers provided in the technology questionnaire.

We supplemented the interview questions by asking the participants to hand-draw sketches of their digital eco-system on paper.
Having participants sketch as part of interviews, has its roots in Cultural Probes \cite{Gaver1999Design:Probes} and has been used successfully within the Human Computer Interaction (HCI) and Computer Supported Cooperative Work (CSCW) communities in a number of study contexts. For example ~\cite{Kaye2014MoneyFinances} asked participants to draw a map of their finances to understand how people track money. Building on work by e.g., \citet{Ryan2009DeviceArtifacts} on mapping interactive digital artefacts, \citet{vertesi_data_2016-1} used the data narrative approach to promote people's descriptions of how they manage their personal data. Drawing was used to facilitate more thorough conversations, to elicit grounded comments about data practices and examine conceptions of personal data space. \citet{vertesi_data_2016-1} argue that drawing during interview allows the remembering of new stories, the discovery of forgotten elements, and the visual expression of relationships between devices and data. The drawings produced are not a test of a participant’s precision or recall with regard to their digital ecosystem, rather, they give structure to the interview process as tools with which to think \cite{vertesi_data_2016-1}. Memory aspects aside, asking participants to engage with their personal information sharing in this way rather than, for example, simply talking about it or using their devices as the main support, exploits the power of defamiliarization described in work by e.g., \cite{Bell2005MakingStrange} to invite participants to have a new perspective on familiar aspects of their lives.

In the current work, participants were asked to come to the session prepared with pen/pencil and paper. They were asked to sketch out their communications channels, devices, types of information shared/intended recipient(s), along with whether usage was in a personal or professional capacity, and (separately) whether they were using a personal or professional account. The process of drawing the sketches proceeded throughout the interviews with participants adding to them as elements were remembered. Participants were also asked to identify and draw any links between media, devices and the public or private aspects of shared information.  Participants were later asked to photograph and email the final sketched maps to the interviewer (21 sketches were returned from 26 participants). We have included two example sketches in Appendix C to illustrate the variation in approach and reflecting the uniqueness of each individual in the study and the personal nature of their digital ecosystem.

We made audio recordings with participant consent, and coded anonymised transcripts by performing thematic analysis ~\cite{Braun2006UsingPsychology}, using NVivo. Our analysis took a hybrid approach (see also \cite{Sas2017DesignUsers}): existing concepts were used for deductive coding while new concepts grounded on the empirical data from the interviews, contributed to the inductive coding. The deductive coding included, for example, concepts from technology law literature on personal information such as Right to Erasure and pseudonymous posting \cite {Rosen2011TheForgotten} and on the participants' desires or requirements for a tool to manage their digital traces. The resulting coding list was iteratively refined in the light of the interview data, as new codes emerged. One author did the early coding, undertaking frequent code review sessions with another author to help remove potential biases. The list of preliminary codes was then distilled further into a set of refined codes (Table 1 in Results) that corresponded with a high number of instances across the transcripts, or that captured novel emerging design ideas or relevant practices. All authors were involved in three data sharing meetings and arrived at the designations of the refined codes through discussion. Further iterative analysis and clustering resulted in the three final themes: \textit{visibility and self-disclosure}, \textit{unintentional information leakage} and \textit{digital privacy literacies}. From an original list of six themes, \textit{visibility and self-disclosure} arose from the combination of themes of visibility, self-disclosure and identity curation, \textit{unintentional information leakage} was one of the original themes, while themes of conceptions of overview of own data and data concerns were combined into one as \textit{digital privacy literacies}. Each theme is discussed in turn below illustrated with pseudonymous quotes.

%% file: results.tex
\section{RESULTS}\label{sec_results}
We now discuss each of our themes in turn (see Table 1 for details of the codes that comprise the themes), illustrating points with verbatim quotes. While participants' ages are reported, all names have been changed and other identifying information omitted to protect interviewees’ privacy.

\begin{table}[b]
  \caption{Code Book: themes developed from the interview data with details of the codes that comprise them}
  \label{tab:code_book}
  \begin{small}
  \begin{tabular}{llcc}
    \toprule
    Theme & Code & No. Interviews & Occurrences\\
    \midrule
    \multirow{7}{*}{\textbf{Visibility and Self-disclosure}} & Creating/maintaining a professional online profile & 10 & 18 \\ 
     & Curating a non-professional online identity & 5 & 13 \\
    & Life Events motivating online sharing & 20 & 43\\
    & World Events motivating online sharing & 19 & 40 \\
    & Revealing one’s location & 10 & 22 \\
    & Pseudonymous posting & 6 & 7 \\
    & Context missing/partial picture & 5 & 5 \\
    \midrule
    \multirow{6}{*}{\textbf{Unintentional Info Leakage}} & Revealing too much about yourself & 8 & 10 \\
    & Things posted by others about you & 11 & 17 \\
    & Others revealing more than they intended to & 17 & 23 \\
    & Leakage in domestic settings/IoT & 4 & 4 \\
    & Leakage due to work/personal boundary blurring & 2 & 4 \\
    & Postings appropriated to other media e.g., broadcast & 5 & 7 \\
    \midrule
    \multirow{7}{*}{\textbf{Digital Privacy Literacies}} & Google-searching self & 14 & 14 \\
    & One’s online info being boring/of no interest to others & 9 & 20 \\
    & Lack of agency/overwhelmed/resignation to fate & 4 & 5 \\
    & Concerns about marketing & 10 & 14 \\
    & Inappropriateness & 6 & 9 \\
    & Fake news & 7 & 9\\
    & Security measures taken & 6 & 10 \\
    & Right to erasure & 15 & 21\\
 \bottomrule
\end{tabular}
\end{small}
\end{table}

\subsection{Visibility and Self-disclosure}
\subsubsection{Curating an online profile}
Creation or maintenance of online identities was mentioned by many of our participants: ten spoke about creating or maintaining professional online profiles while five spoke about their curation of a non-professional online identity. Attitudes to having a presence and being in/visible online had an age dimension. On the whole, younger participants (18-25 years) were those who were more likely to speak about the vital importance of having an online presence especially in enhancing, even enabling one’s career. Xander, a recent graduate now working for a student organisation, had a good understanding of how his online information was received and judged by others; he had even been researching how sharing personal information online comprised a form of self-commodification:
\begin{quote}
\textit{To be successful, whether it’s on social media, just to get likes, or whether it’s in a professional sense... you are your brand and you need to sort of, I don’t know, show that in every single way you do online, and so I am just careful about what I post, whereas I know that some other people aren’t, and it can look quite bad.} (Xander, 22)
\end{quote}
Meanwhile, five participants expressed awareness that online traces only ever provided a partial picture, albeit an authentic representation of someone, and as such invited interpretation. Una, a clerical assistant commented: 
\begin{quote}
\textit{I think it’s a true snippet of who I am. But I think that it is just a snippet, because I don’t post about everything…..I wouldn’t be concerned. I think people make judgements no matter what you put out there, so they have a snippet of what it is!} (Una, 21)
\end{quote} 
While not expressing particular concern, Una touched on the potential for inferences to be drawn from incomplete traces, and also on how some interests lend themselves more to being documented and shared, saying: "I play a musical instrument, but that is nowhere on my social media, and not many people know how musical I am... [but] they know I’m maybe quite sporty, but they don’t see the musical side" (Una, 21). As part of his youth worker role, Calvin recruits others to work with the young people for whom he has safeguarding responsibilities on on whom he carries out informal background checks online --- while also mindful of the limitations of these:
\begin{quote}\textit{One of the first things I do is a basic Facebook search of them...and sometimes that has… created a false narrative ... We had a student on placement, and her photos showed ...this quite ragey, party person. When I met her, they were completely lovely, and those photos didn’t represent them fairly.} (Calvin, 29)
\end{quote}
One participant spoke from bitter personal experience about times when the visibility of information posted online by them had directly led to their prospects being seriously curtailed: Tom, a bakery supervisor, had been aggressively confronted at interview with examples of jokes he had posted online. This left him feeling ``humiliated'' and his job interview drew to an abrupt close (Tom, 20). Meanwhile Yan, a civil servant spoke of: 

\begin{quote} \textit{...colleagues who had disciplinary hearings...you have to be cautious...on what you put out online, so if somebody shares a post.. with… swearing and questionable humour, that gets around the office….somebody might be having a quiet word to someone.} (Yan, 23)
\end{quote} 
One interviewee reflected on living with an ongoing tension between feeling they had to promote their visibility online, and protecting their own safety and that of their community. Zara, a third sector worker whose family had sought asylum in the UK, commented that her family’s desire to be invisible online raised questions, including around their legal status, which she perceived as damaging her current prospects:
\begin{quote}\textit{I have also met loads of people who were genuinely risking... running for their lives. And any information that they put online digitally would be instantly sought out, so they stayed off any kind of digital, social media, anything. But then they’re also met with the contrast of needing to put something out in order to progress --- I’m going to say in a Western country --- but, well, that’s not exclusive to Westerners, but just to put yourself on show, or otherwise people don’t think you’re legitimate.} (Zara, 20)
\end{quote}
Helen, a helpline operator, described how wanting to stay offline had affected a loan application: "I had to put myself back on the public [voting] register, because if you use credit at all, you can’t get it if you’re not on the public register... you get... forced into actually having your name and address out there" (Helen, 56).
These experiences demonstrate how a visible online presence can become skewed to tell a particular story. They also give insights into how online privacy can be a privilege, when some individuals must establish and maintain their online representation as a source of public identification ---  for evaluation and validation by others.

\subsubsection{Motives for Self-disclosure} 
As well as talking about the ways in which they disclosed personal information online, participants also described what motivated them to do so. 20 participants described how life events motivated them to increase the amount of information they posted and 19 participants described being motivated to respond to a world event or cause by posting or responding to posts online. These actions revealed online what were otherwise private aspects of themselves. 

Regarding special life events, HR manager Anna’s (38) desire to lose weight prior to her wedding involved tracking steps and recording her meals to calculate calorie consumption on her FitBit. In interview she discussed being unconcerned about the privacy implications, progressing to the subscription version of the FitBit app to record her menstrual cycle. In contrast, Xander, expressed alarm when a friend posted their university acceptance letter on social media --- publicly disclosing an application number, leading Xander to imagine this provoking:  "someone sort of malevolent on the other end of the computer that just wants to mess with people" (Xander, 22). 

With regard to causes,  when probed, two participants who had claimed to share little about themselves online recalled making public contributions to petitions and fundraising pages, and, also, sharing their full name and location, information that often persists online indefinitely. Queenie, a lab technician commented in interview: "I’ve signed petitions… judging by emails that come in unsolicited, you realise that other people know you’re an animal lover" (Queenie, 55). 
Xander had unwittingly revealed his address through location tracking after getting involved in a good cause: "The sports club that I’m involved with did... a charity fundraiser, so I downloaded it to... track my runs... you realise that the start and finish point is right outside your front door" (Xander, 22).
Meanwhile Patrick, an engineer was cognisant that causes he got involved in online and related affiliations and opinions would persist with possible repercussions: \begin{quote} \textit{I have some views which are commonly held and not as yet controversial --- but I do wonder at some point politically if they will become (so). I’m becoming more and more careful about what I post about things like that.} (Patrick, 56)\end{quote} 

It is worth noting here that it was younger participants (18-25 years) who most vividly described their public expressions of support for certain causes  in terms of obligation. Research assistant Rita had clearly tussled with expressing anything other than her more usual apolitical online self: "I thought that I was doing a very, very bad thing, because I was being very silent on the matter [of the Black Lives Matter (BLM) campaign]" (Rita, 25).  Rita consequently contributed to the campaign in her own way by promoting online book lists she had curated about racial injustice.

Participant responses often linked self-disclosure with wanting to share a particular personal event or to signal allegiance with or opposition to a more universally significant concern. Participants recognised the risks of oversharing, but also, as in posting in support of a social movement such as BLM, of under-sharing --- for fear of being seen as uncaring or unaware. Visibility and self-disclosure were perceived as tricky to navigate: even when preferring to maintain a minimal online profile, current norms around recruitment, immigration applications and the financial service industry, for example, often require personal information to be available online in order to assess credibility or eligibility. Participants often implicitly recognised the need to tread a fine line regarding the volume and type of information to put out there. We heard from the perspectives of both recruiters and the recruited, how the absence of certain types of information threatened employment and wider social and professional prospects. This partial picture was recognised as potentially damaging in specific contexts such as job seeking. 

\subsection{Unintentional Information Leakage}
Various forms of self-disclosure can stem from information  leakage. Participants mostly talked about unintentional information leakage in respect of their personal (i.e., non-work) information sharing, but leakage due to work/personal boundary blurring was also reported by two participants. Third sector workers Zara and Flora had both resorted to using their personal Facebook logins to set up Facebook accounts for work, having tried and failed to set up dedicated work profiles. This led to their mixing information sources across professional and private accounts, and their professional identities encroaching onto the personal: "It’s really difficult to keep a personal Facebook, I think, if your workplace is using Facebook a lot for the way they work" (Flora, 54).

Four interviewees reflected on their experiences of information leakage in their domestic lives. Such leakage could be intimate, if also unintentional and/or somewhat creepy, though some were more relaxed about this than others. Will (24) said he was unconcerned about any potential implications of sharing his passwords with his girlfriend. Meanwhile, teacher Linzi (31) expressed her unease at her realisation that she had been inadvertently sharing details of her plans and contacts with her partner via her calendar, due to device synching. On reflection, she said that while she had nothing to hide, in different circumstances --- such as in relationships involving domestic violence and controlling behaviours --- this could have been hugely problematic. Another interviewee (Queenie, 55) expressed obvious discomfort during interview when recalling her discovery that her partner could view her Internet searches, including those she had conducted to check particular spellings. 

A further instance of information leakage in a domestic setting was described by designer Ben, who recalled instances of inadvertent cross-device leakage.  His Smart TV would spontaneously display his neighbour’s private smart phone information: "the TV switched itself on... someone [in the adjoining property] was playing on their phone or something... and if they’ve clicked a Share button... it sometimes comes through to my TV in the middle of a show" (Ben, 49). These information leakages mostly occurred due to insecure account settings. However, 11 interviewees recounted how things posted online about them by others either unthinkingly or, due to more deliberately malicious actions, were also sources of unintentional information leakage. Healthcare worker Will described how his mother's misplaced pride had potentially serious consequences: "I said (to her) I wanted to join… one of the intelligence agencies. She… posted up, `My son wants to become a spy, how does he do it?'" (Will, 24). While Will joked, saying this was "a little bit counter-productive” for undercover work, he went on to recount how his mother had also posted details of his confidential military achievements and training exercises on Facebook, including the nature and location of the training. This leakage violated military protocols and created a security risk.

Instances of postings either by participants or people known to them, that had then been appropriated to other media, were reported by five participants. Conversations believed to be private had been recorded and/or more widely (re)shared in a very different, typically much later context, e.g.,:
\begin{quote}
\textit{A friend of mine… was called out on Twitter for things that he’d said in a private group chat six years ago, and…reported to his place of study. It’s all cleared up, but it still sort of hit home... people can take anything you say and change it to however they see it… I guess if there’s any shred of doubt, it can be... disastrous.}(Xander, 22)
\end{quote}
Delivery driver Vinny (24) found one of his old Tweets  embedded in a newspaper article several years after the article's publication. He had concerns about assumptions people might make about his beliefs when seeing the tweet out of context: "it was talking about...outdated cultural stereotypes...and I was wondering why my name was involved in this [article he’d found]" (Vinny 24).

In contrast to more narrow understandings of data leakage that link it to the use of insufficiently secure settings, our participants reported threats that arose in many other ways. Such threats were apparent through the actions of others, via the re-appropriation of content intended for a specific audience, into new locations and contexts. This revealed the information to a wider public --- often alongside personally-identifying, or socio-politically significant information. This might be done with the intent of causing reputational damage, but even in the absence of malicious intent, such combinations of digital traces can lead to revelations, and the associated lack of control that individuals have can be problematic and/or distressing. Our participants' responses also show that the proliferation of Internet of Things (IoT) devices has created new vectors for information leakage. Personal and specifically domestic contexts are where they reported the majority of such experiences, perhaps due in part to the Lockdown context, but also because these instances were those that were the most immediate, identifiable and relatable. Two female participants in particular talked of recently discovering routes via which partners had sight of what had previously been private information, and of their raised awareness, and associated concern.

\subsection{Digital Privacy Literacies}
The interviews were revealing with regard to participants' digital privacy literacy, their knowledge and understanding of the nature and potential of coherent digital traces, and of their control and personal agency over those traces. Overwhelmingly, participants were aware that there was a great deal of publicly available information online about them. 14 participants explicitly mentioned that they knew this from having previously conducted a Google search on themselves. Jenny, a local government officer, had been disturbed to discover some information related to a company with whom she'd long ago been involved: "it had my full name and address and date of birth, and I was like, 'Whoah!'" (Jenny, 43) Despite this, none of those who reported having searched for themselves on Google was able to describe with confidence or accuracy the range of publicly or semi-publicly (for instance, a Facebook account) visible information across their various online accounts. For those participants who said they had never Googled themselves, when offered the opportunity to check what was visible online during the interview, they were invariably surprised by the level of detail about them that was public. Further, slighty more than half (14) of the participants said they had what they regarded as stringent approaches to information sharing, deletion and account security. For example Olly, an electronic engineer, recounted one of his practices and the motivation behind it: "I tend to disable search history...my bigger fear is I am an immigrant into this country…so I think that if I search for something because it was in the news, can it be connected to me...by mistake?" (Olly, 40). However, this was at odds with some of participants' other answers during interview. For example, when asked how they would advise a friend with the same digital services as them what security measures should be taken to secure their information, in the case of all but 6 participants, their answers tended to be very general and provided few specific recommendations, indicating that they had relatively low levels of privacy literacy and a lack of awareness around their own information’s potential for being compromised.   

As referenced in the wider results, participants did, on the whole, have a good sense of the long-lasting persistence of their information, once it was online. However, envisaging the potential effects of connecting apparently discrete aspects of their online information coherently proved more challenging even with access to their various accounts and having their sketches to hand. Understandably then, participants struggled to comment on where potential risks or possible consequences lay where others might connect the same dots into coherent digital traces to potentially use against them. Where matters of concern were expressed, these related to interviewees being aware of inviting unwelcome marketing (10 participants), and being targeted by advertised goods, particularly if these were of no personal interest to them. 

We found it striking that nine of the participants regarded themselves and the personal information they had shared online as being of no likely interest to others. Ivor, a writer and teacher, was aware that a great deal of his personal information was available but was bemused that there could be any interest in it: 
\begin{quote}
\textit{They would know where I live. They know who I know, they know who I interact with a lot, they know what I work as, they know what my interests are... They could easily work out any political, spiritual, other viewpoints. They would know what I don’t like, they would know what I do like, and not just in terms of products. As I say, I’m quite a boring individual, so none of it would be particularly shocking, and it would struggle to get a PG [UK film classification Parental Guidance: for children 8+] certificate, but at the back of my mind, it’s 'why do you want to know all this? Why do you want to harvest all this about everyone and everything?'} (Ivor, 59)
\end{quote}
This self-identification as "boring", a term used specifically by five participants, was a factor in some interviewees’ lack of motivation around deleting superfluous personal information circulating online. Four further participants used language such as "dull" or "uninteresting" to describe their postings.

In line with the literature, four participants expressed a lack of agency, overwhelm or resignation ~\cite{Barth2017TheReview,Hargittai2016WhatApathy}, or feeling unable to manage and where necessary remove information ~\cite{Obar2020SunlightAssistance}. Flora, for example, conveyed a sense of being overwhelmed by her online clutter or, at least, with finding sufficient time to deal with it; the task was clearly not a priority: "I would take one look at it and go, 'Oh, my God!' and walk away: I’ve got far too much to do to untangle this mess" (Flora, 54). Helen meanwhile, mentioned her sense of resignation to fate when asked to make a suggestion for a designed tool or service that could help. She found it difficult to conceptualise and dismissed the notion as she would be unlikely to use one: "I'm not sure what I'd want from it. I think in some ways, I’ve kind of accepted that the horse has bolted!" (Helen, 56). Another participant shared a compelling first hand experience to support their thinking that any remedial action could only ever be of limited use, and the benefits were unlikely to outweigh the substantial detriment of living in online obscurity:
\begin{quote}
\textit{I did have a stalker... I sort of removed everything that I used to have [online], and I just existed in the dark for a long time. But it wasn’t dark enough, because she did actually come to [my workplace]... so at that point I realised there was just no point in being dark on all of these tools, because ....if they want to find you, they will.} (Matt, 44) \end{quote}

Participants across all age groups understood how time and a potentially changed future context of a posting affected its significance and 15 participants talked about their right to erase such digital traces. Notably, those in the 18-25 years age group were more likely than older participants to express a desire to delete the digital traces of their younger selves, as these no longer represented the person they perceived themselves to be. Xander, 22 had recently deleted everything on Facebook posted in the years pre-university, while Zara, 20 had increased her privacy settings to prevent others from seeing her childhood postings. As Vinny, a delivery driver explained of the online material most concerning him: "I’d probably delete the stuff from most of my school days... I’ve forgotten a lot of the stuff I posted back then, but... some of it might be potentially embarrassing" (Vinny 24).

In summary, participants showed an awareness that they had left digital traces but could not be accurate about how visible their online information was: those who looked were surprised when confronted with the reality. Even with the support of sketching, the majority did not or could not show how their devices and information channels were interconnected. All participants had a sense of the persistence of online information but only rarely did they acknowledge the potential for connections to be made and compromises to arise. Even those who believed themselves vigilant in their approach to personal information sharing could not explain beyond the most basic guidance, how one would secure devices and channels to minimise risks from cumulative revelation. In general, participants lacked agency to undertake remedial action to their digital traces out of a sense of it being too late, not possible to do, or it not being necessary as their traces were "boring" and therefore posed little risk to them. Younger participants were more likely to have actively removed content, often to delete elements of childhood online activity as they moved into adulthood.

%% file: conclusions.tex
\section{DISCUSSION AND CONCLUSION}\label{sec_conclusions}

This research focused on employed people’s everyday online information sharing practices and their associated levels of awareness of how pieces of personal information --- their digital traces --- can interconnect over different online channels and media over time. We wanted to find out to what extent people recognised how these connected traces are available to others, to be explored as a more coherent whole; what this coherent whole could convey about them, including their apparently private self (e.g., behaviours, values, routines etc.); and where and to what extent they were aware of hazards and potential harms of how this could be used against them, and by whom. Through thematic analysis of the outcomes of 26 interviews, we uncovered themes of \textit{visibility and self-disclosure}, \textit{unintentional information leakage}, and \textit{digital privacy literacies}. 

\textit{Visibility and self-disclosure} was heavily influenced by necessity and obligation, with some participants feeling compelled to have an online presence when job seeking. This is consistent with Berkelaar and Buzzanell's findings \cite[p84]{Berkelaar2015OnlineCapital} that employers increasingly expect potential staff to maintain digital career capital to enable employers to ``construct and evaluate professional and/or workplace identities''. Participants also identified that online visibility could help to build legitimacy as citizens, and to comply with perceived social norms --- for example, by publicly expressing a stance around current events such as Black Lives Matter. Choosing \textit{not} to be visible and \textit{not} to disclose information about oneself can be seen as a privilege, afforded to those who are established members of society and not seeking work. 
This is especially pertinent for those whose safety may be jeopardised by online visibility --- such as survivors of domestic or other abuse, and asylum seekers --- yet who feel compelled to be visible due to the adverse impact that invisibility could have on their chances of getting a job, or gaining legitimacy as citizens and members of social groups \cite{Coles-Kemp2020TooSociety}; also see \cite{Strohmayer2019TechnologiesLines}.

\textit{Unintentional information leakage} occurred as a result of the actions of others, who shared participants’ information to unforeseen audiences and at times causing un/intentional shame or other harm. This could be particularly uncomfortable when the information shared was from long-forgotten posts, or was taken out of context.  While participants had a good understanding of the persistent nature of aspects of their traces, they found it difficult to recall what they had previously posted across multiple channels and hence where potential vulnerabilities might lie. Yet it is not at all surprising that participants struggled to remember past posts when remembering involves ``cognitive processing of knowledge from the past, through a repetitive process of reconstruction'' \cite[p371]{vandenHoven2014AExperiences}. Given the volume of information commonly shared online, remembering everything that one has posted presents an intractable cognitive processing burden that links to our third theme of \textit{digital privacy literacy}. While processes of remembering can be supported by a range of internal and external cues including those that are technological such as Facebook ``On This Day'' reminders, things always get irretrievably forgotten \cite{vandenHoven2014AExperiences}. Although options exist to have content removed from the Internet or at least not show up in search results --- e.g., the Right To Erasure \footnote{UK Information Commissioner's Office (ICO, 2017: 49)} --- there is no easy mechanism through which to erase \textit{aspects} of one’s past history online, or to remove comments made over years that could be misinterpreted or show one in a bad light if later taken out of context. Multiple respondents rationalised that their online information was ``boring'' and of no interest to others. They also referred to being unable to summon the required time, effort and/or practical digital privacy competencies to erase aspects of their past history online. This is understandable, when such curation increasingly involves sophisticated multidisciplinary skills and knowledge spanning digital, technical, legal, and socio-cultural competencies. For example, an individual who is seeking public election might want to check back through their past history online for any information that could be taken out of context and wrongly interpreted as expressing socially unacceptable views. Of course, this is also open to misuse, with individuals who genuinely hold socially unacceptable views cleaning up their online profiles to obfuscate their true opinions. 

Our study was conducted at a time when the legal context of the UK was an uncertain one. Materially, little has changed for UK citizens with regard to GDPR since the UK left the EU, as the UK enacted the UK-GDPR in 2020. However there is an ongoing high-profile political and legislative debate as to whether the UK should diverge more aggressively from the European framework, which is seen in some quarters as unnecessarily burdensome and overprotective \cite{2021Data:GOV.UK}. All these changes took place, and were publicly discussed, while the interviews took place. This posed some legal and ethical challenges for the research: knowing that the GDPR would cease to be applicable shortly after the interviews were completed, what legal assurance could be given to the participants? It raised also questions for the substantive part of the research: discussions surrounding post-Brexit data protection in the news will have created more awareness of data protection questions, and may also have contributed to an even stronger feeling of uncertainty and vulnerability.  Disclosing data about oneself in the UK during 2019, the year prior to this study, also meant that it was at least not certain what legal protection would apply to it in a few months' time, which given the permanence of digital traces poses a significant difficulty. 

\subsection{Limitations and Future Work}

A key limitation of our work is that while our participants were able to conceptualise aspects of the implications of personal information sharing in interview, they consistently struggled to conceptualise the entirety or whole picture of their accumulated digital traces across multiple channels and across time, and potential knock-on effects and risks. We acknowledge that the data narrative approach was not sufficient to achieve this and in this context, we identify the following pressing future work:

\begin{itemize}

\item {Demonstrate to individuals in everyday terms --- perhaps by using other narrative approaches, including scenarios --- the potential use by another agent of seemingly harmless pieces of personal information posted across disconnected digital traces.} 

\item {Go beyond ``awareness nudges'' by promoting reflection \textit{before} sharing, to enable people to  make informed choices about the information that they add to their cumulative digital traces.}

\item {Some participants conveyed anxiety around their old posts being re-discovered, despite them not having clear recall of their contents, amplifying their perceived impotence and lack of knowledge about how to go about removing offending information. We see an opportunity to enable people to efficiently curate the material that they have posted in the past online, without having to trawl through every single post or delete an account wholesale.} 

\item {A further opportunity to reduce unintended information leakage lies in integrating prompted password change as part of standard installation for owners of domestic internet-enabled/ IoT devices, combined with information about inherent risks of devices and how to mitigate against these, to protect domestic privacy.}

\end{itemize}

Last but not least, there is a critical need to address digital privacy literacy, including digital privacy gaps. This aligns with ongoing efforts towards ensuring social justice within the HCI e.g., \cite{Strohmayer2019TechnologiesLines}, and CSCW \cite{Talhouk2020FoodInsecurity} design and wider research communities. Talhouk et al.'s CSCW work \cite{Talhouk2020FoodInsecurity} around the digitisation of food aid intended for Syrian refugees in Lebanon found that due to refugees’ ``low technological literacies'', their ``experiences of engaging with food aid'' were severely impoverished and ability to “identify and report the misuse of the technologies by other stakeholders and intermediaries” (p133) curtailed, amplifying the already present power asymmetries experienced by those groups.

While interviewees had a good understanding of the persistent nature of their traces, they found it difficult during interviews to recall what they had previously posted. Remembering, from the Latin --- \textit{rememorari} --- or ``call to mind'', involves cognitive processing of knowledge from the past, ``through a repetitive process of reconstruction'' \cite[p371]{vandenHoven2014AExperiences}. Yet people face challenges in making connections between something visible, and its meaning --- and, to  paraphrase Rancière \cite{2007ArtIndisciplinarity}--- across heterogeneous spaces and times. Rancière also refers to ``the power of art in its ability to represent what is absent or unrepresentable --- and that when they are represented they infer power'' (ibid). He, along with others, discusses memories as works of fiction --- reconstructions as opposed to re-presentations or later reproductions. While processes of remembering can be supported by a range of internal and external cues --- things always get irretrievably forgotten \cite{vandenHoven2014AExperiences}. There is only ever a partial picture, or re-presentation.

Our work offers understandings around personal information and what it collectively comprises \cite{Hogan2015TangibleRelations}, including the inferences that others can draw. It aims to promote personal agency around management of this information. As future work, we will use the findings reported here to inform the design of an online tool. The digital user interface of this tool, as well as how it curates, contextualises, and relates information to people, will be informed by the qualitative outcomes of design workshops. The tool will allow people to explore the risks and consequences surrounding their own online data-sharing activities and the digital traces they leave behind. We are mindful of Elsden et al.'s proposition that ``design should seek to support people in making account of their data, and guard against the assumption that more, or ``better'', data will be able to do this for them''\cite{Elsden2016AInformatics}. Our current design work is also mindful of so-called ``moral economies'' that are produced as a result of practices and activities around personal data, which are laden with affect, cultural expectation, and responsibility~\cite{vertesi_data_2016-1}. Design, we argue, is central to promoting sense-making and digital privacy literacies in this context. Even the provision of designed tools is a form of design activism, introducing new frictions into online activity to provide context to, reflection on and guidance for, our information-sharing decisions \cite{Ash2018DigitalTransition,Gordon2020MeaningfulInefficiencies}, and/or where necessary, helping to develop counter-narratives ~\cite{fuad2013design}. 
